\documentclass[reprint,
superscriptaddress,
groupedaddress,
showpacs,preprintnumbers,
nobibnotes,
amsmath,amssymb,
aps,
pra,
floatfix,
]{revtex4-1}

\usepackage{graphicx}
\usepackage{dcolumn}
\usepackage{bm}
\usepackage{epstopdf}
\usepackage{float}
\usepackage[caption = false]{subfig}
\usepackage{array}
\usepackage{graphicx}
\usepackage{cancel}
\usepackage{ulem}

\begin{document}


\title{ Semi-classical Computation of Low-energy Collisions Cross Section in Mutual Neutralization of Helium and Hydrogen Ions }

\author{Sifiso M. Nkambule}
\email{snkambule@uniswa.sz}

\author{Sibusiso Zwane}%
%


\affiliation{Department of Physics,University of Swaziland, Kwaluseni, M201, Swaziland.}

\date{\today}

\begin{abstract}
A Semi-classical model is employed to theoretically study mutual neutralization in the collisions of $\mathrm{^4He^+}$ and $\mathrm{H^-}$. The model includes nine covalent states of $\mathrm{^2\Sigma^+}$ symmetry. Here the assumption is that only two states are interacting at a given internuclear position. The reaction is studied for collisions energies below 100 $eV$. The total mutual neutralisation cross section is computed and compared with previous results.  
\end{abstract}

\maketitle


\section{Introduction}
 Mutual Neutralization (MN) is a process where positive and negative ions collide, resulting in charge transfer and the formation of neutral atoms. When the MN reaction rate is very large, Whitten \textit{et al} concluded that it can be an important escape channel for the formation of the excimer species~\cite{Whitten82}. When a helium cation and a hydrogen anion collide, the process is
 \begin{equation}
\label{HE}
\mathrm{He}^{-}+\mathrm{H}^{+}\rightarrow\mathrm{He}+\mathrm{H}.
\end{equation}

The process is of interests in many fields of research, such as the chemistry of the interstellar medium~\cite{Glover03} and the gas evolution of the early universe~\cite{Galli98,Galli12}. Helium and hydrogen ions played a crucial role in the formation of important species like $\mathrm{HeH^{+}}$, $\mathrm{He_{2}^{+}}$ and $\mathrm{H_2}$\cite{Galli12}. At the divetor section of a fusion reactor, like ITER~\cite{Iter}, ions of helium and hydrogen are proposed to be present~\cite{Fantz06,Fantz08,Fantz11}. Thus it is of importance to study all the possible reactions, involving the ions, that may take place, including MN.
With the advent of cold ion storage rings, like Desiree~\cite{Desiree} and merged beam facilities, the MN reaction can be studied experimentally.

This MN reaction is good enough for testing theory, due to the size of the species involved. A fully quantum study for this reaction has been recently reported~\cite{Larson16}, where eleven $^2\Sigma^+$ states were included in modelling the nuclear dynamics using the log-derivative~\cite{Johnson85} method. In this study autoinozation amongst the coupled electronic states was also studied. However, autoinization was found to have a very low contribution to the MN total cross section, for energies below 10 $eV$. The results showed a large cross section, comparable with previous results~\cite{Olamba,Peart,Chibisov,Ermolaev92}. The current study includes ten adiabatic states and the aim is to test the reliability of the Landau-Zener~\cite{Landau,Zener} model for the HeH system, since it is not computationally demanding as the log-derivative method.

In Section~\ref{com} of this paper, details on the computations of the tranformation matrix and diabatic potential energy curves are discussed. The results and conclusion are given in section~\ref{res} and section~\ref{conc}, respectivley.

\section{\label{com}Computations}
The MN reaction can be theoretically studied by viewing the ionic and covalent interaction of the potential energy curves. In the adiabatic picture~\cite{Mead82}, the potential energy curves for a diatomic system do not cross~\cite{Stenrup06,Mead82}. Such curves, however, do not preserve the ionic/covalent character of the states. Thus a potential energy curve may exhibit an ion-pair state character at short internuclear distances and a covalent state character at large distances. On the other hand, if the potential energy curves are transformed to a diabatic representation, the character can be preserved. Such potential energy curves, though, will cross each other, even for a diatomic system.

The Landau-Zener~\cite{Landau,Zener} model assumes only two states are interacting at an avoided crossing. The adiabatic potential energy curves obtained by some of us previously~\cite{Larson16}, are used in this study. The adiabatic-to-diabatic transfromation matrix, $\mathbf{T}$ is of the form
\begin{equation}
\mathbf{T}=\left(\begin{matrix}
\cos[\gamma(R)]&\sin[\gamma(R)]\\
-\sin[\gamma(R)]&\cos[\gamma(R)]\\ 
\end{matrix}\right).
\label{strict_2}
\end{equation}

\begin{figure}[!h]
\includegraphics[scale=0.3,angle=-90]{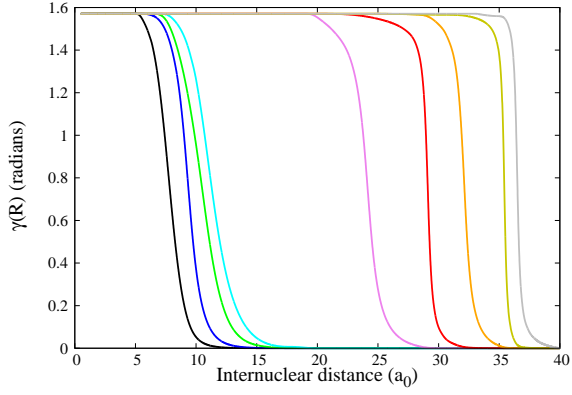}
\caption{\label{coups1} Rotational angles obtained by eq.~\ref{angle_eq} for the potential energy curves of states of the HeH system.}
\end{figure}

\begin{figure}[!h]
\includegraphics[scale=0.3,angle=-90]{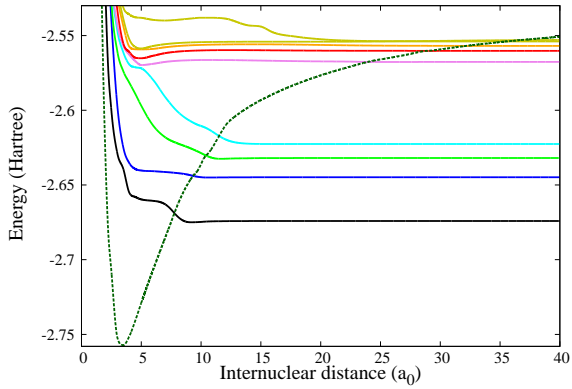}
\caption{\label{dia1} Diabatic potential energy curves for the HeH system.}
\end{figure}

The rotational angle, $\gamma(R)$ is obtained from integrating the first derivative coupling element~\cite{Mead82} obtained by Larson \textit{et al}~\cite{Larson16},
\begin{equation}
\gamma(R)=\int_R^\infty F_{ij}(R')dR',
\label{angle_eq}
\end{equation}

where $F_{ij}(R)$ is the first derivetive non-adiabatic coupling element between states $i$ and $j$. The couplings are at large internuclear distances and are known to drive the MN reaction in many systems~\cite{Nkambule15,Nkambule16,Hedberg14,Stenrup06}. The coupling elements are peaked at the avoided crossing. Thus the rotational angle exhibit a drop by a factor of $\frac{\pi}{2}$ at the avoided crossing. This drop has been observed previously for other systems~\cite{Hedberg14,Nkambule15}. The rotational angles for the HeH system are shown in fig.~\ref{coups1}.
The diabatic potential energy curves (shown in fig.~\ref{dia1}) are assumed to vary linearly with the internucler distance $R$  in the vicinity of the crossing $R_x$, ie
\begin{equation}
V_1(R)-V_2(R)=cR,
\end{equation}
where $V_1(R)$ and  $V_2(R)$ are the diabatic potential energy curves for the two states and $c$ is a constant. The probablity $\mathrm{p_{\ell}}$ for remain in a diabatic curve is gven by~\cite{Landau,Zener}
\begin{equation}
p_{\ell}=\exp\left(\frac{\eta}{V_x}\right),
\end{equation}
where $\eta=\frac{2\pi H_{12}^2}{c}$ and $H_{12}$ is the electronic coupling element. $V_x$ is the radial velocity at the curve crossing. As pointed out previously~\cite{Nkambule15,Hedberg14}, the electronic couplings play a crucial role on the quality of the results one may obtain by using the Landau-Zener method. Here we are using electronic coupling elements obtained from taking the values of the diabatic potentials at the curve crossings.

In the $\mathrm{HeH}$ system, the ion-pair state crosses nine covalent states. The probality for ending in the highest covalent state (here the covalent states are numbered 1-9, starting from the highest in energy, refer to fig.~\ref{dia1}) is given by

\begin{eqnarray}
\wp(ion,1)= \begin{cases}2p_{\ell_{ion}}p_{\ell_{9}}p_{\ell_{8}}p_{\ell_{7}}p_{\ell_{6}}p_{\ell_{5}}\\
\times p_{\ell_{4}}p_{\ell_{3}} p_{\ell_{2}}p_{\ell_{1}}(1-p_{\ell_{1}}),& \text{if } \ell < \ell_1\\
 0, & \text{otherwise}. \end{cases}.
\end{eqnarray}
The Landau-Zener probabilty for a transition from the ion-pair state to a covalent state $n$ is
\begin{eqnarray}
\wp(ion,n)= \begin{cases}p_{\ell_{ion}}p_{\ell_{9}}\cdots p_{\ell_{n}}(1-p_{\ell_{n}}),& \text{if } \ell_{n-1}< \ell < \ell_n\\
 p_{\ell_{ion}}\cdots p_{\ell_{n}}(p_{\ell_{n}})^2(1-p_{\ell_{n}})+\\
 p_{\ell_{ion}}\cdots p_{\ell_{n}}(1-p_{\ell_{n}})^2\\ \times (1-p_{\ell_{n}}),& \text{if } \ell_{n-2}< \ell < \ell_{n-1}\\
 \vdots\\
 p_{\ell_{ion}}\cdots p_{\ell_{n}}(p_{\ell_{n-1}})^2\cdots \\ \times (p_{\ell_{2}})^2(1-p_{\ell_{n}})+
 p_{\ell_{ion}}\\ \times \cdots p_{\ell_{n}}(1-p_{\ell_{n-1}})^2 \cdots\\ \times (p_{\ell_{3}})^2(1-p_{\ell_{2}})^2(1-p_{\ell_{n}}),& \text{if } \ell_{1}< \ell < \ell_{2}\\
 
 p_{\ell_{ion}}\cdots p_{\ell_{n}}(p_{\ell_{n-1}})^2\cdots\\ \times (p_{\ell_{1}})^2(1-p_{\ell_{n}})+\\
 p_{\ell_{ion}}\cdots p_{\ell_{n}}(1-p_{\ell_{n-1}})^2 \cdots \\ \times (p_{\ell_{2}})^2(1-p_{\ell_{1}})^2(1-p_{\ell_{n}}),& \text{if } \ell < \ell_{1}.\\
  \end{cases}
\end{eqnarray}
Here $\ell_{n}$ denotes the maximum rotational quantum number ($\ell_{max}$) attainable before $R_x$ is reached for state $n$;
\begin{equation}
\ell_{max}=2R_x\sqrt{\mu (E+\Delta E)},
\end{equation}
 where $E+\Delta E$ is the total energy of the system, without including the centrifugal barrier term, and $\mu$ is the reduced mass for the species.
 
The values where the ion-pair curve crosses the covalent states, $R_x$, are displayed in table~\ref{tab1}. The electronic couplings are obtained using the ATD method reported in ref.~\cite{Hedberg14}.
\begin{table}
\begin{tabular}{c|cc}
\hline 
 & $R_x$ ($a_0$) & $H_{12}$ ($eV$) \\ 
\hline 
1 & 7.74 & 1.2817$\times 10^{-2}$ \\ 

2 & 9.28 & 3.7956$\times 10^{-3}$ \\ 
 
3 & 10.38 & 6.3394$\times 10^{-3}$ \\ 

4 & 11.51 & 7.8453$\times 10^{-4}$ \\ 

5 & 23.89 & 6.9397$\times 10^{-4}$ \\ 
 
6 & 28.88 & 1.9948$\times 10^{-4}$ \\ 

7 & 32.15 & 2.4096$\times 10^{-4}$ \\ 

8 & 35.39 & 1.5679$\times 10^{-4}$ \\ 

9 & 36.52 & 1.5246$\times 10^{-4}$ \\ 
\hline 
\end{tabular}
\caption{\label{tab1} Crossing distances and values of electronic couplings for the  diabatic covalent states of the HeH system used in the study.} 
\end{table}
The total cross section formula, for state is given by
\begin{equation}
\sigma_n(E)=\dfrac{\pi}{k_{n}^{2}}\sum_{\ell=0}^{\ell_{max}}(2\ell +1)\wp(ion,n),
\label{cr1}
\end{equation}
where $E$  is the collision energy, if we assume the threshold energy to be zero. $k_{n}$ is the asymptotoc wave number of the incoming channel,
\begin{equation}
k_n=\sqrt{2\mu(E-E_{n}^{th}},
\label{cr2}
\end{equation} 

and $E_{n}^{th}$ is the asymptotoc energy of state $n$.
\section{\label{res}Results}
The total MN cross section for collisions of $\mathrm{^4He^++H^-}$ is computed using eq.(~\ref{cr1}) for each of the nine states. The total cross section is then computed using
\begin{equation}
\sigma_{total}(E)=\sum_{n=1}^{9} \sigma_n(E).
\end{equation}
\begin{figure}[!h]
\includegraphics[scale=0.33,angle=-90]{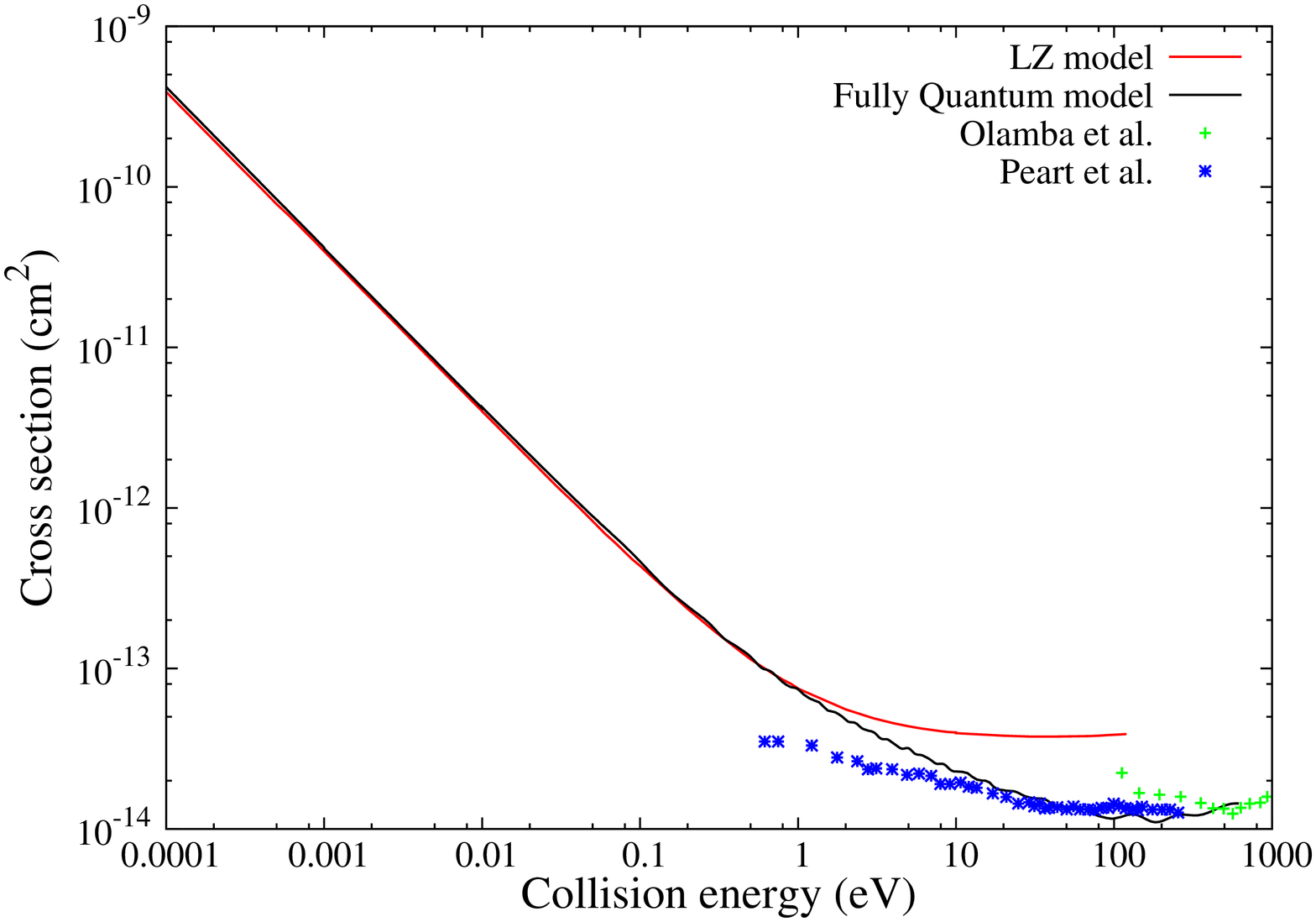}
\caption{\label{cross1} Total cross section for MN of $\mathrm{^4He^++H^-}$ compared with other results~\cite{Larson16,Olamba,Peart}.}
\end{figure}

The MN total cross section results are shown in fig~\ref{cross1}. Results from the current calculations are labelled ``LZ model". Here, they are compared with results from a fully quantum model~\cite{Nkambule16} and some experimental results by Peart \textit{et al}~\cite{Peart} and Olamba \textit{et al}~\cite{Olamba}. The cross ection from the Landau-Zener model is comparable with the fully quantum model at low collision energies. For energies above 1 $eV$, the total cross section from the current model is larger. This is a phenomenon previously observed in other systems~\cite{Nkambule15,Nkambule16}.

\section{\label{conc}Conclusion}

The MN reaction total cross section computed from the Landau-Zener model is comparable with other results for the $\mathrm{^4He^++H^-}$ reraction. At low collision energies the cross section follows the Wigner threshhold law~\cite{Wigner48}. The MN reaction is driven by non-adiabtic couplings at large internucler distances.
\begin{acknowledgements}
We would like to acknowledge Prof. \AA sa Larson and Prof. Ann Orel for providing us with the data for non-adiabatic couplings and the adiabatic potential energy curves used in this study.
\end{acknowledgements}
\normalem
\bibliography{apssamp}
\end{document}